\documentclass[12pt]{iopart}
\usepackage{graphicx}

\begin{document}

\title[Julia set describes quantum tunneling in the presence of chaos]
{Julia set describes quantum tunneling in the presence of chaos}

\author{A. Shudo$^1$,  Y. Ishii$^2$ and K.S. Ikeda$^3$}

\address{1 Department of Physics, Tokyo Metropolitan University, 
Minami-Ohsawa, Hachioji, Tokyo 192-0397, Japan}

\address{2 Department of Mathematics, Kyushu University, Hakozaki, 
Fukuoka 812-8581, Japan}

\address{3 Faculty of Science and Engineering,
Ritsumeikan University, 1916 Noji-cho, Kusatsu 525-0055,  Japan}



\begin{abstract}

We find that characteristics of quantum tunneling in the presence of chaos 
can be regarded as a manifestation of the Julia set of the complex dynamical system. 
Several numerical evidences for the standard map together with a rigorous statement for 
the H\'enon map are presented demonstrating that the complex classical paths 
which contribute to the semiclassical propagator are dense in the 
Julia set.  Chaotic tunneling can thus be characterized by 
the transitivity of the dynamics and high density of the 
trajectories on the Julia set. 

\end{abstract}

\def\C{{\cal C}}
\def\T{{\cal T}}
\def\M{{\cal M}}
\def\Q{{\cal Q}}
\def\S{{\cal S}}
\def\P{{\bf P}}
\def\im{{\rm Im}}
\def\re{{\rm Re}}

Recent studies on tunneling in multidimensions 
have revealed that the existence of chaos 
affects the signature of quantum tunneling. 
The observation of purely quantum mechanical calculation in 
chaotic systems shows that tunneling can become 
{\it chaotic} or chaos seems to {\it assist} tunneling 
\cite{Boh1,Boh2,Tom,Doron,Creagh1,Creagh2,Creagh3,TK1,TK2,SI1,SI2}.  
The idea capturing such a novel aspect of tunneling looks very attractive, 
 but a direct connection between chaos and tunneling can only be 
accomplished by interpreting the quantum phenomenon by the trajectory 
of the classical dynamics \cite{Creagh2,Creagh3,TK1,SI1,SI2}.   
When one is particularly interested in the tunneling process, the use of 
complex trajectories is essential since the transition due to 
tunneling occurs where the real-valued classical trajectories cannot reach. 

The most well-known technique using the complex space is so-called  
instanton method in which the tunneling penetration is evaluated mainly 
by a single classical path moving on the reversed potential \cite{instanton1,instanton2}. 
On the other hand, in chaotic systems, it has been 
found in the time-domain semiclassical analysis 
that a bunch of complex paths almost equally contribute to 
the tunneling transition between classically forbidden 
regions \cite{SI1,SI2}. 
They typically form a tree-like fractal structure in the complex initial 
value plane, 
and its outstanding appearance compels us to prepare some concept which 
controls dominating complex paths in the semiclassical sum of 
contributing candidates \cite{SI1,SI2}. 
All the characteristic structures appearing  
in the tunneling wavefunction in chaotic systems 
originate from it. 

However, the {\it Laputa chain}, which was so introduced 
in \cite{SI1,SI2}, has been still phenomenological so far, 
and remains even mysterious if no link to some concept compatible with  the 
dynamical system theory is made. 
One may thus naturally ask why such a structure play a special 
role in the complex trajectory description of chaotic tunneling, and 
what sort of mechanism underlies such conspicuous objects
in the complex plane. 
The purpose of this Letter is to provide a clear answer to these questions. 
Our final claim is simple and would be rather natural: {\it Julia set is 
the origin of chaotic tunneling}.

Let us begin with introducing the model system we are concerned 
with.  The system we study here is a family of two-dimensional 
area preserving maps, in which 
the mixed phase space being realized in a certain range of the parameter space. 
The time evolution of the phase point $(p, \theta)$ is given as the mapping rule as 
\begin{eqnarray}\label{map}
(p_{n+1},\theta_{n+1}) = 
F(p_n, \theta_n) 
\equiv (H'(p_n) -
V'(\theta_n), \theta_n + H'(p_n) -V'(\theta_n)). 
\end{eqnarray}
\noindent
Here, 
$H(p) = p^2/2$ and $V(\theta) = K\sin \theta$ are the most standard choice, but 
suitable modification or replacement of the kinetic or the potential term 
is sometimes helpful and will be made 
according to the target of the analysis. 

Since the map model does not have the energy as the 
Hamiltonian flow problem does so, one cannot consider 
the tunneling through the energetic barrier, which may 
be a normal setting of the tunneling problem. 
Instead, dynamical confinement due to classically 
disconnected components such as KAM tori and 
chaotic components in the phase space plays the role of barriers, 
and the quantum transition between such invariant regions 
is regarded as tunneling \cite{DavisHeller}.  

At least in the first setting of the problem,  it is not at all obvious that 
several different situations, such as the tunneling transition 
out of the quasiperiodic region into some chaotic component, 
or its reverse process, or that between 
different chaotic components,  
could be treated on the same footing.  
However, as will be described below and also become 
one of the most important point in the present report, 
the choice of initial and final states does not matter to 
the whole story.

Typical quantum mechanical wavefunctions in the mixed 
phase space are displayed in Fig. 1.  
In both models,  the tails of the wavefunctions do not monotonically 
decay even in the tunneling regime, rather there appear several 
unexpected structures; the crossovers 
of the slope, the plateau regions and irregular interference patterns
on it.  All these characteristics are only qualitatively featured 
\cite{SI1,SI2}, but 
they are commonly observed not only in the dynamical tunneling 
problem, but also in the energetic barrier tunneling \cite{OSIT}.  

The semiclassical approach, which is extensively developed 
in recent studies of quantum chaos or quantum chaology \cite{Gutzwiller}, 
 works quite well even when one employs it as a tool 
describing the tunneling process.  Apart from an added technical 
(but sometimes crucial) problem originating from the Stokes phenomenon
\cite{SI3}, 
which we do not enter into details here,  our task in the 
semiclassical analysis is essentially the same as the real one, that is, 
to evaluate the Van-Vleck propagator:
\begin{eqnarray}\label{prop}
\Psi_n(p_0, p_n) \approx
\sum_{\stackrel{\scriptstyle p_0=\alpha}
{p_n=\beta}}
A_{n}(p_0, \theta_0)
\exp \left\{ -\frac{i}{\hbar} S_{n}(p_0, \theta_0)\right\},
\end{eqnarray}
\noindent
where the summation is taken for all $(p_0,\theta_0)$ which satisfy 
the boundary conditions for the initial momentum $p_0=\alpha$ and
the final momentum $p_n=\beta$.
Here, $S_{n}(p_0, \theta_0) =\sum_{j=1}^{n}
[H(\theta_j)-V(\theta_j)+\theta_j(p_j - p_{j+1})]$
is the action along a classical trajectory, and
$A_{n}(p_0, \theta_0) 
= [2\pi \hbar (\partial p_n/\partial \theta_0)_{p_0}]^{-\frac{1}{2}}$
represents the amplitude factor
associated with its stability.


\begin{figure}[h]
\begin{center}
\includegraphics[width=.80\linewidth]{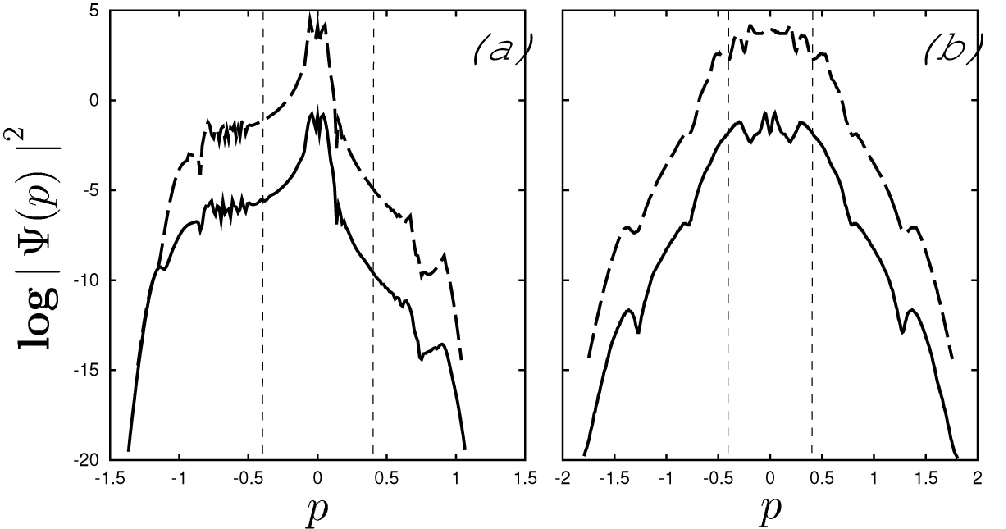}
\end{center}
\caption{
Quantum(bold line) and semiclassical wavefunction
(dashed line) for (a) the model 
with $H_0(p) = \frac{p^2}{2}\frac{(p/p_d)^6}{(p/p_d)^6+1}  + \omega p$, 
and $V(\theta )  = K\sin \theta$, where $p_d = 5$, $\omega = 2$ and $K= 1.2$, 
and (b) the model with $H_0(p) = \frac{p^2}{2}$ and $V(\theta )  
= K\sin \theta$, where $K = 1.5$. In both cases, 
the initial wavepacket is set as $\Psi(p) = \delta (p)$.  
The real-valued classical orbits cannot reach the region outside the dashed lines
within the time step taken here ($n = 6$ for (a) and $n=5$ for (b)). 
The semiclassical wavefunction is shifted in order to clarify the structure. 
}
\label{Fig:FIG1}
\end{figure}

Since we here take the $p$-representation,
$p_0$ should be a real quantity. So, the canonical partner $\theta_0$ may
be used to identify the (complexified) trajectories
contributing to the sum (\ref{prop}), and it is then allowed to be complex as
$\theta_0=\xi+i\eta\ \ (\xi,\eta$ real).
We visualize the contributing complex paths by displaying the
set~\cite{SI1,SI2}:
\begin{eqnarray}\label{mset}
\M_n \equiv \bigcup_{\beta\in {\bf R}}{\cal M}_n^{\ast , \beta}
=\bigcup_{\beta\in {\bf R}}
\{(p, \theta) \in {\bf C}^2 \ | \ p_n=\beta \}
\end{eqnarray}
\noindent
on the $\theta_0$-plane of the slice $\{p_0=\alpha \}$
for some initial condition $\alpha \in {\bf R}$.
The set $\M_n$ on the $\theta_0$-plane, 
which usually looks like clouds or wisteria trellis on
a macroscopic scale, is decomposed into finer and finer
structures as it is magnified \cite{SI1,SI2}. 
One can see that its basic element is a string with various scales.
Each string represents a trajectory appearing 
in the semiclassical sum (\ref{prop}). 
We note that in the integrable limit only the branches
connected with the real plane (i.e., $\eta=0$) survive and all other
complicated objects disappear \cite{SI1,SI2}.

A huge amount of candidate paths may discourage us since it appears to be
no more possible to establish a simple view of tunneling in the presence 
of chaos.  
However, among all possible candidates the complex paths 
forming a sequential structure, which runs 
in the vertical direction at the center of Fig. 2(a) and clearly discernible
from the other aggregated strings, 
exceed any other candidate paths in amplitude.
We have called such a characteristic structure
the {\it Laputa chain} \cite{SI1,SI2}.
As shown in Fig. 1,  one finds that semiclassical sum including 
{\it only} such complex paths contained in the Laputa chains has reproduced
almost all details of tunneling into chaotic regions.
Our task is, therefore, reduced to clarifying what 
this marked structure appearing in the initial value plane represents.


\begin{figure}[h]
\begin{center}
\includegraphics[width=.80\linewidth]{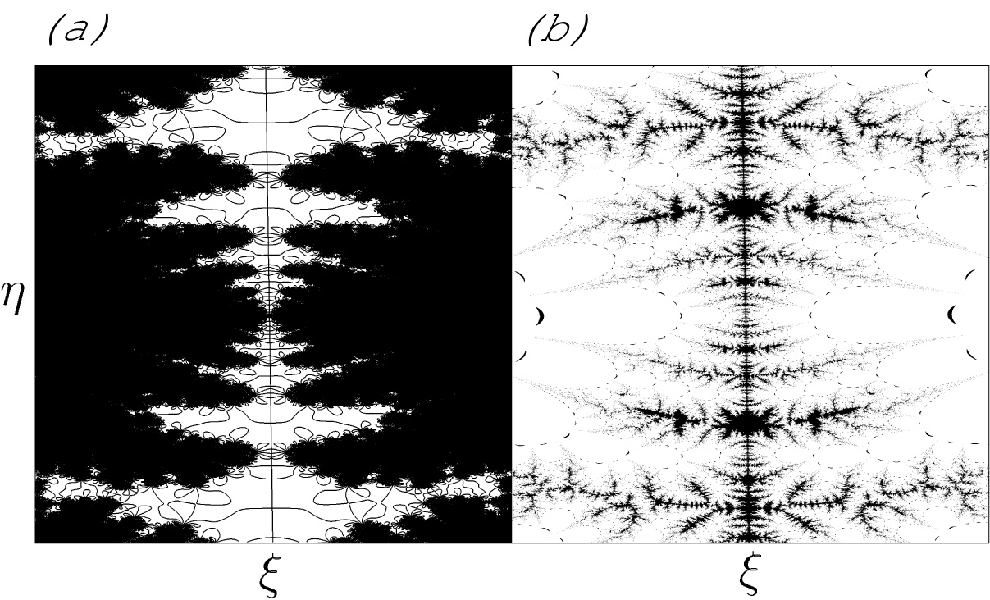}
\end{center}
\caption{
(a) A magnification of the initial value representation $\M_n$ on 
the $\theta_0$-plane in case of the standard map with $K=1.5$ and $n=40$.  
The range shown above is given as $4.197487346\times 10^{-2} \le 
\xi \le 4.197487362\times 10^{-2}$, and $6.693592\times 10^{-4} 
\le \eta \le 6.693601\times 10^{-4}$. 
The initial momentum is set as $p_0=0$. $\M_n$-set consists of a bunch of 
self-similar objects, whose basic element is a string. Except for the 
central part, such 
strings are so densely aggregated that 
individual string cannot be 
resolved in this scale. However, magnifying the black 
    area, one can find a similar structure in the scale shown here, that is, 
    the black area is also composed of a bunch of string objects. 
Each string represents an individual component of
the semiclassical sum (2).
The strings running in the vertical direction look as if they cross with each other,
but actually they avoid with very narrow gaps,
that is, the strings form a serial chain-like structure connected via 
narrow gaps.
(b) The slice of $K^+$ by $\{p_0 = \alpha \}$. 
This is numerically obtained by plotting 
the initial points whose trajectories remain a ball in ${\bf C}^2$ 
with a certain finite radius, $r=10^3$, in this case. 
}
\label{Fig:FIG2}
\end{figure}


The reason why some complex paths dominate the others in the 
semiclassical sum (\ref{prop}) is, in general, that the imaginary part of 
their action, $\im \, S_{n}(p_0,\theta_0)$, are relatively small. This is 
because the absolute value of each term in (\ref{prop}) is 
mainly governed by $\im \, S_{n}(p_0, \theta_0)$, 
rather than the amplitude factor $A_{n}(p_0,\theta_0)$. 
This in turn means that the complex paths forming the Laputa chain 
should gain small imaginary action as compared to the other paths not 
forming the chain structure.  Indeed,  as shown later,  the trajectories 
initially placed on the Laputa chain approach the real $(p,\theta)$-plane 
exponentially, which provide minimal or relatively small imaginary action. 

Conversely, one can say that this property characterizes the Laputa chain 
and makes them distinguishable from the others. 
Furthermore, they are specific in that those trajectories stay 
in bounded regions because after coming close 
to the real plane they almost follow the behavior of the trajectories 
on the real plane and the real orbits  are all bounded in the present situation. 

This is a hint to link the Laputa chain to a proper object compatible 
with the theory of  dynamical systems,  since the Julia set, 
which plays a  central role in the complex dynamical systems, 
is specified as the set satisfying such a property. 
More precisely, the {\it forward Julia set} $J^+$ is defined as the
boundary of the set $K^+$ of points whose forward orbits remain
in a finite region~\cite{JuliaSetDef}:
\begin{equation}
K^+=\{ \ (p,\theta)\ |\  \{F^n(p,\theta)\}_{n>0} \ {\rm is \ bounded \ }\}
\end{equation}
and
\begin{equation}
J^+=\partial K^+.
\end{equation}
The polynomial diffeomorphism like H\'enon map $f$, 
which is defined on ${\bf C}^2$, has a polynomial inverse, 
so both the forward and the backward
iterations can be considered. In such a case 
we define $K^+$ (resp. $K^-$) as the set of points in ${\bf C}^2$
whose forward (resp. backward) orbits are bounded, and $J^{+}$
(resp. $J^-$) to be the boundary of $K^{+}$ (resp. $K^-$)
which we call the {\it forward}
(resp. {\it backward}) {\it Julia set}. The set $J\equiv J^+ \cap J^-$
is called the {\it Julia set} of $f$.
The forward (or backward) Julia set $K^{\pm}$ is where the orbits have sensitive dependence
on initial conditions, which means that the chaotic motion
is realized on it.

Remarkably enough, such a purely mathematical object enters into
physics as quantum tunneling in chaos.
Indeed, as shown in Fig. 2(b),  the similarity between  the
chain-shaped structure demonstrated in Fig. 2(a) and the 
slice of $K^+$  by the same plane is obvious. 
The slice of $K^+$ shows a typical dendrite-like structure which 
often appears in the one-dimensional complex dynamical systems 
\cite{ComplexChaos}.
The number of strings constituting the Laputa chain 
increases in an exponential rate,  so coincidence between them 
becomes better as the time proceeds \cite{SI2,SII}.  Note that 
the orbits put on the highly aggregated branches surrounding the Laputa chain 
do not stay in a finite phase space domain but rapidly 
escape to infinity.  

It is possible to provide a rigorous statement if one focuses
on the cubic potential model given by putting $H(p) = p^2/2$ and
$V(\theta) = c\theta - \theta^3/3$.
The map (\ref{map}) is transformed to a standard form of the
H\'enon map:
\begin{equation}
f: (x, \, y) \longmapsto (y, \, y^2+(1-c)-x),
\end{equation}
by the affine  change of coordinate $(p, \theta )=(y-x, \, y-1)$.
The H\'enon map is known to be one of the simplest nonlinear systems
in the two-dimensional space, and its dynamics is extensively studied
by several authors. Among them, investigation from the
complex dynamical point of view has been developed in the last decade
(see, for example, \cite{BS1,BS2,BS3} and the references therein)
by using the pluripotential theory, the theory of currents, etc.

As in the case of the standard map, 
it is reasonable to focus on the $\im \, S_{n}(p_0,\theta_0)$ of each 
trajectory, but to be compatible with the invariant set of the 
dynamical system one should consider the set of trajectories 
having the property described above in the limit of $n$ going to infinity. 
The most natural condition would be to select the complex orbits whose 
$\im \, S_{n}(p_0,\theta_0)$ has a finite limit even when $n$ goes to infinity. 
Such a filtering only serves as a necessary condition for semiclassically 
contributing orbits, but it is at least true that 
the trajectories whose $\im \, S_{n}(p_0,\theta_0)$  
are divergent cannot contribute to the semiclassical summation 
since those orbits either tends to zero in their magnitude 
or will be removed by the Stokes phenomenon \cite{SI3}. 
Therefore we define the {\it Laputa chains} as,   
\begin{eqnarray}\label{laputachain}
{\cal C}_{\mathrm{Laputa}}\equiv
\bigl\{(p, \theta)\in {\cal M}_{\infty} \bigm | \im
 \,  S_n(p, \theta) \
{\rm converges} \ 
{\rm absolutely \ at} \ (p, \theta) \bigr\}
\end{eqnarray}
In this definition ${\cal M}_{\infty}$ is an object introduced to 
represent the limit of $\M_n$-set when $n$ goes to infinity. 
More precisely, 
\begin{equation}
{\cal M}_{\infty}\equiv
\bigcup_{\beta\in{\bf R}}{\cal M}_{\infty}^{\beta},
\end{equation}
where ${\cal M}_{\infty}^{\beta}$ is given as the Hausdorff limit
of ${{\cal M}^{\ast, \beta}_{n}} \equiv
\bigl\{ (p, \theta)\in {\bf C}^2 \bigm| \, p_n=\beta \bigr\}$
(compare eq. (3)). Thus, the set ${\cal M}_{\infty}$ corresponds to
${\cal M}_n$ for the time step `$n=\infty$'.
It is possible to prove that this Hausdorff limit itself contains 
the forward Julia set $J^+$ \cite{SII,SI2}, which in itself is a partial 
verification of our numerical observation. 
The following assertion concerning 
the relation between ${\cal C}_{\mathrm{Laputa}}$ 
thus defined and $J^+$ is proved by the second-named author:

\vspace{3mm}
\noindent
{\bf Theorem.}
{\it Let $F$ be the time-one map on ${\bf C}^2$ associate to
the kicked rotor (\ref{map}) with $H(p) = p^2/2$ and $V(\theta) = c\theta - \theta^3/3$}, 
{\it and $h_{top}(F)$ be the topological entropy of $F$,}
%
 \begin{enumerate}
 \renewcommand{\labelenumi}{(\roman{enumi})}
  \item {\it If $h_{\mathrm{top}}(F|_{{\bf R}^2})>0$,
        then $\overline{{\cal C}_{\mathrm{Laputa}}}\supset J^+$.}
  \item {\it If $F$ is hyperbolic on $J$ and
        $h_{\mathrm{top}}(F|_{{\bf R}^2})>0$,
     then $\overline{{\cal C}_{\mathrm{Laputa}}} = J^+$.}
  \item {\it If $F$ is hyperbolic on $J$ and
          $h_{\mathrm{top}}(F|_{{\bf R}^2}) = \log 2$,
        then ${\cal C}_{\mathrm{Laputa}} = J^+$.}
 \end{enumerate}
Here $\overline{X}$ indicates the closure of the set $X$.
The rough sketch of the proof is as follows~\cite{SII,SI2}:
that $h_{\mathrm{top}}(F|_{{\bf R}^2})>0$ implies the existence
of a saddle periodic point $X$ in the real phase space.
A principal tool we will employ is the following result which is
established by Bedford and Smillie \cite{BS1,BS2,BS3}:
For a complex one-dimensional locally closed
submanifold $M$ in either $J^+$ or an algebraic variety,
there is a constant $c>0$ so that
\begin{equation}
\lim_{n \to +\infty}\frac{1}{2^n}[f^{-n}M]=c \cdot dd^cG^+
\end{equation}
in the sense of current, where $[M]$ is the current of integration
of $M$, i.e. $[M](\phi)\equiv \int_M\phi|_M$, and $dd^c$ is the complex Laplacian. 
In this statement,
$G^+$ represents the Green function for $K^{+}$ given by
\begin{equation}
G^{+}(x, y)\equiv \lim_{n\to +\infty}\frac{1}{2^n}
\log^+ \left\Vert {f}^{n} (x, y) \right\Vert.
\end{equation}
It is easily shown that the support of $dd^cG^+$ coincides with $J^+$.
From this theorem we see that the stable  manifold of
any periodic saddle $p$ is dense in $J^+$,
that is, $\overline{(W^s(p))} = J^+$. Using this result,
together with the fact that the Hausdorff limit
${\cal M}_{\infty}$ contains $J^+$, we obtain the desired claim.

This claim gives a mathematical verification to the observation 
numerically found.  
Indeed, as shown in Fig. 3(a), the trajectories leaving the Laputa chains 
approach exponentially to the real $(p,\theta)$-plane.  
This makes Im\,$S_{n}(p_0,\theta_0)$ converge absolutely. 
As a demonstration of the theorem, we show in Figs. 3(b) and (c) 
the set $\bigcup_{\beta < \beta_0}{\cal M}_n^{\ast, \beta}$ for 
a fixed $\beta_0$ and the 
slice of $J^+$ by $\{ p_0 = \alpha \}$ for the H\'enon map. 
One can see how $\M_n$-set shrinks to the $J^+$ 
as a function of $n$. 


\begin{figure}[h]
\begin{center}
\includegraphics[width=.80\linewidth]{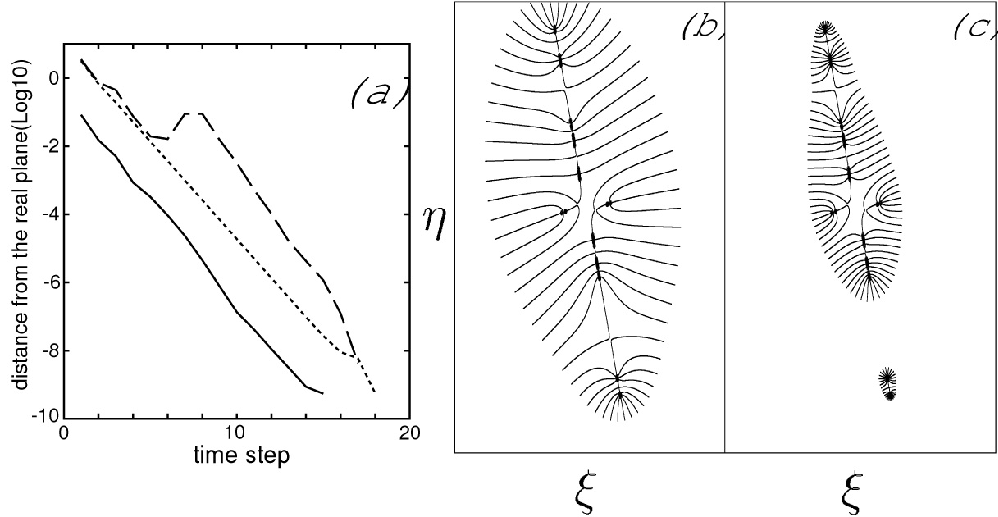}
\end{center}
\caption{
(a) The distance of from the real plane as a function of time is 
displayed for the orbits whose initial conditions are put 
on the Laputa chains.  
The solid line denote the case of the standard map. 
The dotted and broken lines are the ones for the H\'enon map. 
When the Julia set $J$ exists only on the real plane, 
the orbits always approach the real plane directly(dotted line),
 but otherwise the orbits first move around in 
${\bf C}^2$ space, and then approach the real plane (broken line). 
(b)-(c) The set $\bigcup_{\beta < \beta_0}{\cal M}_n^{\ast, \beta}$
($\beta_0 = 10^{10}$)  for the H\'enon map is shown as the solid curves 
in case of (b) $n=9$ and (c) $n=10$. 
The slice of $J^+$ by $\{ p=\alpha \}$ is shown as the dots in each figure. 
}
\label{Fig:FIG3}
\end{figure}


Notice that the assumption $h_{\mathrm{top}}(F|_{{\bf R}^2})>0$ 
in the above theorem is
a mathematical expression which corresponds to the fact that
the underlying classical dynamics $F|_{{\bf R}^2}$ is chaotic.
We also note that the slice of the forward Julia set $J^+$ by
$\{p=\alpha\}$ can be shown to have positive capacity for
any initial condition $\alpha \in {\bf R}$.
Thus, the theorem above suggests that, unlike the instanton
solutions in the integrable case,
a bunch of paths in ${\bf C}^2$ contributes to the tunneling
phenomena if the underlying classical mechanics is chaotic.

It should be noted that the assumption in (i) covers the system 
with mixed phase space, which is the most generic situation in 
physics. In addition, the physical implication or interpretation of another
theorem of Bedford and Smillie on the transitivity of
the dynamics~\cite{BS1,BS2,BS3}
is suggestive in our problem. It states that
for any ${\bf C}^2$-neighborhoods of any two points in the chaotic regions
there is an orbit in ${\bf C}^2$ connecting them,
even in the case where the chaotic regions in the real plane are mutually
disjointed by KAM tori.  This property exactly guarantees the transition 
between any disconnected regions on the  real-valued classical dynamics, 
 and non-zero tunneling amplitude of the wavefunction in arbitrary regions is 
always realized due to the transitivity on the Julia set. 

In this way,  with the help of strong mathematical claims, 
which could be established only by extending the dynamics to the 
complex space, we can clearly understand the reason why chaos seems to
assist tunneling and can become chaotic; these can be attributed to such
{\it high density} of the tunneling paths in $J^+$ and 
the {\it transitivity} of the complexified dynamics.
So far, the structure of the Julia set has been an object which mainly attracts
the interest of mathematicians. But the present result implies that
the Julia set is really observable as chaotic tunneling in various physical and
chemical phenomena. 

\end{document}